\documentclass[11pt]{article}
\usepackage{amsmath,amssymb,epsf}

\def\ben{\begin{equation}}
\def\een{\end{equation}}

  \let\n=\nu

\let\C=\Chi

\def\nn{\nonumber} \def\bd{\begin{document}} \def\ed{\end{document}}
\def\ds{\documentstyle} \let\fr=\frac \let\bl=\bigl \let\br=\bigr
\let\Br=\Bigr \let\Bl=\Bigl
\let\bm=\bibitem
\let\na=\nabla
\let\pa=\partial \let\ov=\overline
\newcommand{\be}{\begin{equation}}
\newcommand{\ee}{\end{equation}}
\def\ba{\begin{array}}
\def\ea{\end{array}}
\def\ft#1#2{{\textstyle{\frac{\scriptstyle #1}{\scriptstyle #2}}}}
\def\fft#1#2{\frac{#1}{#2}}
\def\del{\partial}
\def\vp{\varphi}
\def\sst#1{{\scriptscriptstyle #1}}
\def\oneone{\rlap 1\mkern4mu{\rm l}}
\def\td{\tilde}
\def\wtd{\widetilde}
\def\ie{\rm i.e.\ }
\def\dalemb#1#2{{\vbox{\hrule height .#2pt
        \hbox{\vrule width.#2pt height#1pt \kern#1pt
                \vrule width.#2pt}
        \hrule height.#2pt}}}
\def\square{\mathord{\dalemb{6.8}{7}\hbox{\hskip1pt}}}
\newcommand{\ho}[1]{$\, ^{#1}$}
\newcommand{\hoch}[1]{$\, ^{#1}$}
\newcommand{\bea}{\begin{eqnarray}}
\newcommand{\eea}{\end{eqnarray}}
\newcommand{\ra}{\rightarrow}
\newcommand{\lra}{\longrightarrow}
\newcommand{\Lra}{\Leftrightarrow}
\newcommand{\ap}{\alpha^\prime}
\newcommand{\bp}{\tilde \beta^\prime}
\newcommand{\tr}{{\rm tr} }
\newcommand{\Tr}{{\rm Tr} }
\def\0{{\sst{(0)}}}
\def\1{{\sst{(1)}}}
\def\2{{\sst{(2)}}}
\def\3{{\sst{(3)}}}
\def\4{{\sst{(4)}}}
\def\5{{\sst{(5)}}}
\def\6{{\sst{(6)}}}
\def\7{{\sst{(7)}}}
\def\8{{\sst{(8)}}}
\def\n{{\sst{(n)}}}
\def\cA{{{\cal A}}}
\def\cB{{{\cal B}}}
\def\cF{{{\cal F}}}
\def\tV{\widetilde V}
\def\tW{\widetilde W}
\def\tH{\widetilde H}
\def\tE{\widetilde E}
\def\tF{\widetilde F}
\def\tA{\widetilde A}
\def\im{{{\rm i}}}
\def\tY{{{\wtd Y}}}
\def\ep{{\epsilon}}
\def\vep{{\varepsilon}}
\def\R{\rlap{\rm I}\mkern3mu{\rm R}}
\def\bD{{{\bar D}}}

\def\R{\rlap{\rm I}\mkern3mu{\rm R}}
\def\bD{{{\bar D}}}
\def\R{{{\mathbb R}}}
\def\C{{{\mathbb C}}}
\def\H{{{\mathbb H}}}
\def\CP{{{\mathbb C}{\mathbb P}}}
\def\RP{{{\mathbb R}{\mathbb P}}}
\def\Z{{{\mathbb Z}}}
\def\bA{{{\mathbb A}}}
\def\bB{{{\mathbb B}}}
\def\bC{{{\mathbb C}}}
\def\bD{{{\mathbb D}}}
\def\bE{{{\mathbb E}}}
\def\bZ{{{\mathbb Z}}}
\def\Re{{{\frak{Re}}}}
\def\Im{{{\frak{Im}}}}
\def\cosec{{\,\hbox{cosec}\,}}
\def\Gm{{\Gamma_{\!\! -}}}
\def\Gp{{\Gamma_{\!\! +}}}
\def\stan{{standard }}
\def\nonstan{{supernumerary }}

\begin{document}
\begin{flushright}
MIFP-04-22 \\
{\bf hep-th/0411218}\\
November\  2004
\end{flushright}

\begin{center}

{\large {\bf A Note on Einstein-Sasaki Metrics in $D\ge 7$}}

\vspace{20pt}

W. Chen\hoch{\ddagger},  H. L\"u\hoch{\ddagger 1}, C.N.
Pope\hoch{\ddagger 1} and J.F. V\'azquez-Poritz\hoch{\ast 2}

\vspace{20pt}

\hoch{\ddagger} {\it George P. \&  Cynthia W. Mitchell Institute
for Fundamental Physics,\\
Texas A\&M University, College Station, TX 77843-4242, USA}

\vspace{10pt}

\hoch{\ast} {\it Department of Physics, University of Cincinnati,\\
Cincinnati OH 45221-001, USA}

\vspace{10pt}

\hoch{\ast} {\it School of Natural Sciences, Institute for Advanced Study,\\
Princeton NJ 08540, USA}

\vspace{40pt}

\underline{ABSTRACT}
\end{center}

   In this paper, we obtain new non-singular Einstein-Sasaki spaces in 
dimensions $D\ge 7$.  The local construction involves taking a circle
bundle over a $(D-1)$-dimensional Einstein-K\"ahler metric that is
itself constructed as a complex line bundle over a product of 
Einstein-K\"ahler spaces.  In general the resulting Einstein-Sasaki
spaces are singular, but if parameters in the local solutions satisfy
appropriate rationality conditions, the metrics extend smoothly onto
complete and non-singular compact manifolds.

{\vfill\leftline{}\vfill \vskip 10pt \footnoterule {\footnotesize
\hoch{1} Research supported in part by DOE grant
DE-FG03-95ER40917.

{\footnotesize \hoch{2} Research supported in part by DOE grant
DOE-FG02-84ER-40153.}} \vskip -12pt}  \pagebreak


\newpage

\section{Introduction}

   Einstein metrics admitting Killing spinors are of considerable
interest in string theory and M theory, since they can provide
supersymmetric backgrounds of relevance to the AdS/CFT
correspondence \cite{malda}. For the case of an Einstein-Sasaki
space $X_{2n+3}$, a solution with the geometry AdS$_d\times
X_{2n+3}$ is expected to be dual to a $d-1$ dimensional
superconformal field theory with reduced supersymmetry. Such
solutions arise in the near-horizon limit of certain $p$-branes
located at the tip of the corresponding Calabi-Yau cone
$C(X_{2n+3})$ \cite{dlps,kleb,fig,acharya,morrison}. For example, an
M2-brane on a cone with special holonomy $SU(4)$ interpolates
between AdS$_4\times X_7$ and Minkowski$_3\times C(X_7)$, which
implies that there is an RG-flow in the quantum field theoretical
picture \cite{acharya}.

   Until recently, the known explicit Einstein-Sasaki metrics were
relatively sparse.  Well-known examples are the round sphere in any
odd dimension, and the sphere with the non-standard ``squashed''
Einstein metric in dimensions $D=4n-1$, described as a coset
$Sp(n+1)/Sp(n)$. Other examples include the five-dimensional $T^{1,1}$
space (which is topologically $S^2\times S^3$), and higher-dimensional
analogues.  Aside from these isolated examples, which are all
homogeneous, there were various existence proofs for further
inhomogeneous Einstein-Sasaki metrics, including, for example, 13 on
$S^2\times S^3$ \cite{galit}.  The collection of examples increased
dramatically recently, with the explicit construction of infinitely
many inhomogeneous non-singular Einstein-Sasaki metrics in all odd
dimensions $D=2n+3\ge 5$ \cite{gaunt1,gaunt2}.

    An Einstein-Sasaki metric can always be viewed as a circle bundle
over an Einstein-K\"ahler base space, written as
\be d\hat s^2 = (d\psi' + 2\cA_\1)^2 + ds^2\,,\label{circle} \ee
where $d\cA_\1$ is proportional to the K\"ahler form for $ds^2$.
(See, for example, \cite{gibharpop} for an explicit discussion of this.)
The Einstein-K\"ahler bases $ds^2$ used in \cite{gaunt1,gaunt2}
are the class of such metrics that were constructed in
\cite{berber,pagpop}.  These Einstein-K\"ahler metrics were
themselves obtained as two-dimensional bundles over
Einstein-K\"ahler base metrics $d\td s^2$ of dimension $2n$:
\be
ds^2 = \fft{d\rho^2}{U(\rho)} + \rho^2 U(\rho)\, (d\tau' + \cB_\1)^2
 + \rho^2\, d\td s^2\,,\label{bbpp}
\ee
where $d\cB_\1$ is proportional to the K\"ahler form for $d\td
s^2$.

  The $2n+2$-dimensional Einstein-K\"ahler metrics obtained in
\cite{berber,pagpop} are generally singular. However, this need not
necessarily imply that the Einstein-Sasaki metrics on circle bundles
over them are singular.  Indeed, the main subtlety in the construction
of \cite{gaunt1,gaunt2} consists in showing that the Einstein-Sasaki
metrics can, for suitable choices of parameters, be extended smoothly
onto compact manifolds, even though the Einstein-K\"ahler base spaces
by themselves are singular.\footnote{An analogous approach can be used
to demonstrate that the $G_2$ holonomy metrics constructed in
\cite{clgpg2} as $SU(2)$ bundles over singular self-dual Einstein
4-spaces are also complete and non-singular \cite{yasui}.}

   In this paper, we obtain further examples of non-singular
Einstein-Sasaki metrics, by generalising the construction
described above.  Specifically, we do this by extending the
construction of $2n+2$-dimensional Einstein-K\"ahler metrics to
cases where the $2n$-dimensional base metric is a product of
Einstein-K\"ahler factors $d\td s_i^2$, rather than a single one;
\be
ds^2 = dt^2  + c^2\, (d\tau' + \cB_\1)^2 + \sum_i a_i^2\, d\td s_i^2\,,
\ee
where $c$ and $a_i$ are function of the radial variable $t$.
Although we find that the metrics $ds^2$ are generally singular,
in certain cases the Einstein-Sasaki metric $d\hat s^2$ given by
(\ref{circle}) can extend smoothly onto a non-singular manifold, even 
though the Einstein-K\"ahler base space is singular.

    This paper is organized as follows. In section 2, we give a
detailed exposition of the construction for the case where $ds^2$
is a six-dimensional Einstein-K\"ahler metric constructed as a
two-dimensional bundle over a product $S^2\times S^2$ base. We
obtain seven-dimensional Einstein-Sasaki metrics, which is the
lowest dimensionality for which the generalisation extends beyond
the results in \cite{gaunt2}. In section 3, we generalize the
construction to higher dimensions by using a base space composed
of a product of an arbitrary number of Einstein-K\"ahler spaces,
each of arbitrary even dimensionality. Conclusions are presented in
section 4.

\section{Seven-Dimensional Einstein-Sasaki Metrics}

\subsection{The six-dimensional Einstein-K\"ahler base}\label{ek6sec}

   We begin by constructing six-dimensional Einstein-K\"ahler metrics
of the form
\be
ds_6^2 = dt^2 + c^2\,  (d\tau' + \cB_\1)^2 + a^2 d\Omega_2^2 +
b^2 d\td \Omega_2^2\,,\label{6met}
\ee
where
\be
d\Omega_2^2 = d\theta^2 + \sin^2\theta d\phi^2\,,\qquad
d\td \Omega_2^2 = d\td \theta^2 + \sin^2\td\theta d\td\phi^2\,,
\ee
and the connection $\cB_\1$ is such that
\be
d\cB_\1= p\, \Omega_\2 +q\, \td\Omega_\2\equiv
-p\, dA_\1 - q\,d\td A_\2\,,
\ee
where $\Omega_\2$ and $\td\Omega_\2$ are the volume forms of the two
unit 2-spheres.  We shall take
\be
A_\1=\cos\theta\, d\phi\,,\qquad \td A_\1 = \cos\td\theta\, d\td\phi
\,.
\ee
In order that the circle bundle over $S^2\times S^2$ be well defined,
the ratio $p/q$ must be rational, so that the periods dictated for
$\tau'$ by the consideration of the bundle over each $S^2$ factor are
commensurate.  By a rescaling of $\tau$, one can then, without loss of
generality, choose $p$ and $q$ to be relatively-prime integers.  They
characterise the winding numbers of the circle bundle over the two
2-spheres of the base. Without loss of generality, $p$ and $q$ can be
taken to be positive.  In fact, when we construct the Einstein-Sasaki
7-metrics as circle bundles over these 6-metrics, it will turn out
that the ratio $p/q$ no longer needs to be rational.

   To impose the K\"ahler condition on (\ref{6met}), we begin by choosing
a complex structure for which the K\"ahler 2-form is
\be
J = c\,dt\wedge (d\tau' + \cB_\1) + a^2 \, \Omega_\2 +
b^2\, \td\Omega_\2\,.\label{kform}
\ee
A necessary condition for K\"ahlerity is then that $dJ=0$, implying
\be
\fft{\dot a}{a}=\fft{p\, c}{2a^2}\,,\qquad
\fft{\dot b}{b}=\fft{q\, c}{2b^2}\,.\label{con1}
\ee
In fact, in this case it is easily established that this already
implies that $J$ is covariantly constant, $\nabla_k J_{ij}=0$, and
thus (\ref{con1}) constitutes necessary and sufficient conditions
for (\ref{6met}) to be K\"ahler.

  Now, we impose the additional requirement that (\ref{6met})
be an Einstein metric.  We shall choose the normalisation $R_{\mu\nu}=5g^2\,
g_{\mu\nu}$.  Calculating the Ricci tensor for (\ref{6met}), we
find that the Einstein condition implies
\bea
-\fft{\ddot a}{a}  - \fft{\dot a^2}{a^2} -\fft{2\dot a\,\dot
b}{a\,b} - \fft{\dot a\, \dot c}{a\,c}
-\fft{p^2\, c^2}{2a^4} + \fft{1}{a^2} &=& 5 g^2\,,\nn\\
-\fft{\ddot b}{b}  - \fft{\dot b^2}{b^2} -\fft{2\dot a\,\dot
b}{a\,b} - \fft{\dot b\, \dot c}{b\,c}
-\fft{q^2\, c^2}{2b^4} + \fft{1}{b^2} &=& 5 g^2\,,\nn\\
-\fft{\ddot c}{c} -\fft{2\dot a\,\dot c}{a\,c} - \fft{2\dot b\,
\dot c}{b\,c}
+\fft{p^2\, c^2}{2a^4} + \fft{q^2\,c^2}{2b^4} &=& 5 g^2\,,\nn\\
-\fft{2 \ddot a}{a} - \fft{2\ddot b}{b} -\fft{\ddot c}{c} &=&
5g^2\,.\label{d6eom} \eea
Substituting (\ref{con1}) into (\ref{d6eom}), we find that
\be
\fft{\dot c}{c} = - \fft{p\, c}{2a^2} - \fft{q\,c}{2b^2} +
\fft{1-5g^2\,a^2}{p\,c}\,,\label{ek1}
\ee
together with an algebraic constraint
\be
p - q + 5g^2 (q\, a^2 - p\, b^2)=0\,.\label{ek2}
\ee

   Summarising our results so far, we have shown that (\ref{6met})
is an Einstein-K\"ahler metric if the following system of equations is
satisfied:
\bea
&&\fft{\dot a}{a}=\fft{p\, c}{2a^2}\,,\qquad \fft{\dot
b}{b}=\fft{q\, c}{2b^2}\,,\qquad \fft{\dot c}{c} = - \fft{p\,
c}{2a^2} - \fft{q\,c}{2b^2} +
\fft{1-5g^2\,a^2}{p\,c}\,,\nn\\
&&p - q + 5g^2 (q\, a^2 - p\, b^2)=0\,.\label{ek6}
\eea

   To solve (\ref{ek6}), we introduce a new radial variable $r$ such that
$dr=c\,dt$, leading straightforwardly to
\bea
a^2 &=& p\,r + \ell_1\,,\qquad
b^2 = q\,r + \ell_2\,,\nn\\
c^2&=& \fft{2}{a^2\,b^2} \int_0^r
a^2\,b^2\,\Bigl(\fft{1-5g^2 a^2}{p}\Bigr)\,
dr'\nn\\
&=& -\fft{r}{12 p\, (p\, r+ \ell_1)(q\, r+ \ell_2)}\,
\Big[15 g^2\, p^2\, q\, r^3 + 4p\, [5g^2\, (p\, \ell_2 + 2 q\, \ell_1 -q)]\,
r^2
\nn\\
&& + 6[5 g^2 \ell_1\, (q\, \ell_1 + 2 p\, \ell_2) - (q\, \ell_1 + p\, \ell_2)]
\, r + 60 g^2\, \ell_1^2\, \ell_2 - 12\ell_1\, \ell_2\Big]\,.\label{abcsol}
\eea
The algebraic constraint in (\ref{ek6}) becomes
\be
p-q + 5g^2\,  (q\ell_1 - p\ell_2)=0\,.\label{algcon}
\ee
Note that the choice of origin for $r$ is arbitrary, since $r$ does
not appear explicitly in the equations.  We have exploited this by
choosing the lower limit of the integration in (\ref{abcsol}) to be at
$r=0$.  This implies that the two integration constants $\ell_1$ and
$\ell_2$ in (\ref{abcsol}) are non-trivial parameters.

\subsection{The seven-dimensional Einstein-Sasaki metrics}

    Having obtained the six-dimensional Einstein-K\"ahler base metrics
$ds^2$, we can now proceed to the construction, via (\ref{circle}), of
the seven-dimensional Einstein-Sasaki metrics.  With the K\"ahler form
$J$ for $ds^2$ given by (\ref{kform}), we can introduce the following
potential $\cA_\1$, such that $J=d\cA_\1$:
\be
\cA_\1=r\,d\tau' - a^2 \, A_\1 - b^2 \, \td A_\1\,.
\ee
The Einstein-Sasaki 7-metric is then given by
\be
d\hat s_7^2 = k^2\,  (d\psi' + 2\cA_\1)^2 + ds_6^2\,,\label{es7}
\ee
where the Einstein condition implies that we must have
\be
k^2 = \ft58\, g^2\,.
\ee
The Ricci tensor of $d\hat s^2$ then satisfies $\hat R_{ab} =
\ft{15}{4}g^2\, \hat g_{ab}$.  It is convenient to choose a
normalisation such that $\hat R_{ab}= 6 \hat g_{ab}$, implying
that $g^2=8/5$. Hence, $k=1$ and the six-dimensional
Einstein-K\"ahler metric satisfies $R_{ij} = 8 g_{ij}$.

    The six-dimensional Einstein-K\"ahler metrics $ds_6^2$ that we
obtained in section \ref{ek6sec} generally do not extend smoothly
onto complete non-singular manifolds. We take the radial
coordinate $r$ to range between two zeros of the metric function
$c(r)$, $r_-\le r\le r_+$, for which $a(r)$ and $b(r)$ remain
non-vanishing. In order for the metric to have a smooth extension,
$c(r)$ must approach zero at the two endpoints at the appropriate
rate.  This rate determines the period required for $\psi$ in
order that the metric in the $(r,\psi)$ plane extend smoothly onto
$\R^2$ at the ``origin'' $r=r_-$ or $r=r_+$.  In order for the
metric to extend globally onto a smooth manifold, the periods for
$\psi$ at the two endpoints need to be identical, and must be
consistent with that allowed by the requirement of
well-definedness of the 1-form $(d\tau' + \cB_\1)$.  These
multiple criteria are, in fact, not fulfilled for the
six-dimensional Einstein-K\"ahler metrics $ds_6^2$.

    Nevertheless, as we mentioned earlier, this does not necessarily
imply that the seven-dimensional Einstein-Sasaki metric $d\hat s_7^2$
on the circle bundle over $ds_6^2$ is singular.  We therefore need to
study the global structure of $d\hat s_7^2$ carefully, using
techniques of the kind described in \cite{gaunt1,gaunt2}.  We find
that it is appropriate to define new fibre coordinates $\tau$ and
$\psi$, related to $\tau'$ and $\psi'$ by
\be
\psi' = 2\tau\,,\qquad \tau' = \fft{8\tau -\psi}{8\beta} \,.
\label{taupsi}
\ee
In terms of these, we can re-express the Einstein-Sasaki metric
(\ref{es7}) as
\bea
d\hat s_7^2 &=& \fft{dr^2}{c^2} + \fft{c^2}{16(c^2 + 4(\beta + r)^2)}
(d\psi -A_\1 - \wtd A_\1)^2 + a^2\, d\Omega_2^2 + b^2\, d\wtd\Omega_2^2
\label{regularform}\\
&&\!\!\! + \fft{c^2 + 4(\beta +r)^2}{\beta^2} \Bigl(d\tau - \ell_1 A_\1 -
\ell_2 \wtd A_\1 - \fft{c^2 + 4r(\beta +r)}{8(c^2 + 4(\beta+r)^2)}
(d\psi - A_\1 -\wtd A_\1)\Bigr)^2\,,\nn
\eea
where
\be
\beta = \fft{8\ell_1 - 1}{8p} = \fft{8\ell_2 -1}{8q}\,.\label{betaeq}
\ee
(This equation, which follows from a global consideration, will be 
discussed below.)
Note that (\ref{betaeq}) is consistent with the algebraic constraint
(\ref{algcon}), since we have made the normalisation choice $g^2=8/5$.

      The metric has a rescaling symmetry under which $p\rightarrow
\lambda\, p$, $q\rightarrow \lambda q$ and $r\rightarrow r/\lambda$.  
Thus only the ratio $p/q$ of the parameters $p$ and $q$ is 
non-trivial.  This ratio is determined in terms of $\ell_1$ and
$\ell_2$ by the algebraic constraint (\ref{algcon}), which is
rewritten, after setting $g^2=8/5$,  in (\ref{betaeq}).
It should be emphasised that in the discussion of the six-dimensional 
Einstein-K\"ahler space in section \ref{ek6sec}, $p/q$ had
to be rational in order that the circle bundle over $S^2\times S^2$ 
was non-singular, but that requirement is, as we shall see, no longer 
necessary when considering the regularity of the Einstein-Sasaki space.
The parameters $p$ and $q$ need only satisfy the constraint (\ref{betaeq}).
Thus as far as local considerations are concerned, we have a
family of Einstein-Sasaki metrics described by the two non-trivial
real parameters $\ell_1$ and $\ell_2$.  They characterise the size of
the $S^2$ bolts.

   In our solution (\ref{abcsol}) for the metric functions $a$, $b$
and $c$, we chose an integration constant so that $r=0$ is one of the
zeros of the function $c(r)$.  Thus we take $r$ to lie in the range
$0\le r\le r_+$, where $r_+$ is the smallest positive zero of $c(r)$.
(We can always, without loss of generality, choose to consider $r$
non-negative.)  The functions $a(r)$
and $b(r)$ should remain non-zero in the entire range $0\le r\le r_+$.
We therefore have the following conditions:
\bea
&&\ell_1 >0 \,,\qquad \ell_2 >0\,,\nn\\
&&r_+ >0\,,\qquad c(r_+)=0\,,\qquad (c^2)'(0)>0\,.
\label{l12cons}
\eea
Note that from the last condition, it follows that $c^2(r) >0$ for 
$0<r<r_+$, and that $(c^2)'(r_+)<0$.

     To study the global structure, we first consider the base
manifold, whose metric is given by the terms appearing in the first
line of (\ref{regularform}) (\ie the terms orthogonal to the
$\tau$ fibres).  It can be viewed as an $S^2$ bundle over
$S^2\times S^2$, where the $S^2$ bundle is coordinatised by $r$
and $\psi$.  Without
loss of generality, we may assume that $0\le p\le q$, in which case the
positivity of $a^2$ and $b^2$, together with the conditions
(\ref{l12cons}), implies that
\be
\ft18 (1-\fft{p}{q}) < \ell_1 <\ft18\,,\qquad
0<\ell_2 <\ft18\,,\label{l12limit}
\ee

    The Killing vector $\del/\del\psi$ degenerates at the points
$r=0$ and $r=r_+$ where $c(r)$ vanishes.  In order for the metric
to extend smoothly onto these points, it is necessary that the
period of $\psi$ be commensurate with the slope of $c^2(r)$.
Specifically, if $c^2(r)$ has slope $(c^2)'(r_0)=K(r_0)$ at one of these
endpoints, say $r=r_0$, then writing $c^2(r) \sim K(r_0)\, (r-r_0)$
nearby, and defining $\rho^2=r-r_0$, we see from (\ref{regularform})
that in the $(r,\psi)$ frame we shall have
\be
ds^2\sim \fft{4}{K}\, \Big(d\rho^2 + \fft{K^2\, \rho^2}{256 (\beta+r_0)^2}
\, d\psi^2 \Big)\,.
\ee
This implies that $\psi$ must have period
\be
\Delta\psi = \Big|\fft{32\pi (\beta+ r_0)}{K(r_0)}\Big|\,.\label{psiperiod}
\ee
The periods determined by these conditions at $r_0=0$ and $r_0=r_+$
will agree if (see (\ref{abcsol}))
\be
\beta\, \Big( r_+ + \fft{8\ell_1 -1}{8p}\Big) = 
 (\beta+ r_+)\, \Big(\fft{8\ell_1 -1}{8p}\Big)\,,
\ee
which is satisfied if
\be
\beta = \fft{8\ell_1 -1}{8p}\,. 
\ee
This, together with the relation in terms of $\ell_2$ implied by
(\ref{algcon}), gives the conditions appearing in (\ref{betaeq}).
Note that (\ref{psiperiod}) now implies that $\psi$ has period $2\pi$.
\footnote{One might think that different linear coordinate 
transformations from $\tau',\psi')$ to 
$(\tau,\psi)$ could lead to inequivalent results, but it is easy to
show that (\ref{taupsi}) is the unique possibility, up to trivial 
scalings and shiftings.}  

      The $U(1)$ fibre parameterised by the coordinate $\tau$ in
(\ref{regularform}) never collapses, and so it follows that the period of
$\tau$ is governed only by the connection on the fibre, given by
\be d\tau - \ell_1 A_\1 - \ell_2 \wtd A_\1 - \fft{c^2 + 4r(\beta
+r)}{8(c^2 + 4(\beta+r)^2)} (d\psi - A_\1 -\wtd A_\1)\,. \label{dtau}
\ee
The global structure can be examined by looking at all the cycles
at $r=0$ and $r=r_+$ where $c^2$ vanishes.  They are given by
\bea
r=0:&& A_\1: \qquad 2\pi\,\ell_1\,,\qquad\qquad
\wtd A_\1:\qquad 2\pi\,\ell_2\,,\nn\\
r=r_+:&& A_\1:\qquad 2\pi\,(\fft{r_+}{8(\beta + r_+)} -\ell_1)
\,,\qquad
\wtd A_\1:\qquad 2\pi\,(\fft{r_+}{8(\beta + r_+)} -\ell_2)
\,,\nn\\
&&d\psi:\qquad  2\pi\,(\fft{r_+}{8(\beta + r_+)})\,.
\eea
For the expression in (\ref{dtau}) to be globally extendible, the
ratios of the above quantities must all be rational.  Thus there are
two independent requirements, namely
\be
\fft{\ell_1}{\ell_2} = \alpha\equiv \hbox{rational number}\,,\qquad
\fft{r_+}{(\beta+r_+)\ell_1}=\gamma \equiv \hbox{rational number}
\,.\label{rational}
\ee
One then solves the cubic polynomial for $r_+$ that follows from
setting $c(r_+)^2=0$ in (\ref{abcsol}). Using the two
rationality conditions (\ref{rational}), together with (\ref{algcon}), 
enables us to
express $\ell_2$ purely in terms of $\alpha$ and $\gamma$:
\bea 
0 &=& 1536\alpha^2\,\gamma^3\,\ell_2^4+64\alpha\,\gamma^2\,\Big(
\alpha\,(\gamma-96) -30\gamma\Big) \ell_2^3\nn\\ &+& 8\gamma\,\Big(
32\alpha^2\,(36-\gamma)+27\gamma^2+4\alpha\, \gamma\,(72+7\gamma)
\Big) \ell_2^2\nn\\
&+& \Big(
384\alpha^2\,(\gamma-16)-32\alpha\,\gamma\,(24+7\gamma)-\gamma^2\,
(192+29\gamma) \Big)
\ell_2\nn\\ &+& 48\alpha\,(16-\gamma)+16\gamma\,(2\gamma-3)\,.
\label{l2con}
\eea
Appropriate choices of rational values for $\alpha$ and $\gamma$ lead
to a countable infinity of solutions for $\ell_2$, which in general is
real but not necessarily rational, satisfying the condition
$0<\ell_2<1/8$ specified in (\ref{l12limit}).

\subsection{Further remarks}

    The regular Einstein-Sasaki metrics that we have obtained are
parameterised by the two rational numbers $\alpha$ and $\gamma$,
subject only to the condition that $\ell_2$ following from
(\ref{l2con}) satisfy $0<\ell_2 <1/8$.  In the case where
$\ell_1=\ell_2$, the solutions are included within those discussed in
\cite{gaunt2}.  

    Although in general $\ell_2$ need not be, and indeed is not,
rational, special cases can arise where $\ell_2$ {\it is} rational.
Since $\ell_1= \alpha\, \ell_2$, where $\alpha$ must be rational, it
follows from (\ref{betaeq}) that if $\ell_2$ is rational then $p/q$
is rational.  It also follows from (\ref{rational}) that $\beta$, and
hence $r_+$, must then be rational too.
Using the scaling symmetry discussed previously, one 
can then choose $p$ and $q$ to be relatively-prime integers.  In the 
special case with $(p,q)=(1,2)$, the polynomial expression for
$c^2$ in (\ref{abcsol}) factorises, giving
\be
c^2=-\fft{r(r+\ell_2)(128 r^2 + 128 r\ell_2 + 64\ell_2^2 -1)}{
(2r+\ell_2)(16r + 8\ell_2 +1)}\,,
\ee
and hence
\be
r_+=\fft{\sqrt{2-64\ell_2^2}-8\ell_2}{16}\,.
\ee
For $r_+$ to be rational, it is necessary that $\ell$, defined by
$64\ell^2 + 64\ell_2^2=2$, be rational, in which case $r_+$ is
given by
\be
r_+=\ft12 (\ell-\ell_2)\,.
\ee
Thus the existence of a rational solution amounts to find rational
solutions for $64\ell^2 + 64\ell_2^2=2$, in which one of $\ell$ and
$\ell_2$ must be less than $\ft18$, and the other greater than
$\ft18$. Let $\ell_2$ be less than $\ft18$.  Having a rational
solution for $64\ell^2 + 64\ell_2^2=2$ is then equivalent to having
integer-valued solutions to $x^2+y^2=2z^2$.  One can find many integer
solutions, by using a computer enumeration, and presumably there are
infinitely many. Here, we present a few explicit examples:
\bea
(\ell_1,\ell_2) = (\fft{30}{40}, \fft{1}{40} )\,,&&
c^2= \fft{4r(3-40r)(1+10r)(1+40r)}{5 (3 + 40 r) (1 + 80 r)}
\nn\\
(\ell_1,\ell_2) = (\fft{3}{34}, \fft{7}{136} )\,,&&
c^2= \fft{2r(1-17r)(7+136r)(15+136r)}{17(3 + 34 r) (7 + 272 r)}
\,,\nn\\
(\ell_1,\ell_2) = (\fft{5}{52},\fft{7}{104})\,,&&
c^2= \fft{2r(5-104r)(3+26r)(7+104r)}{13 (5 + 52 r) (7 + 208 r)}
\,.
\eea

     For the cases with $p \ne 2q$, the analysis is much more
complicated.  For $(p,q)=(1,3)$, we did not find any rational
solutions.  It is not clear whether such solutions are intrinsically
absent, or whether our search was insufficiently exhaustive.  For some
other values of integer $(p,q)$, we have found isolated rational solutions.

   We should again emphasise. however, that the parameters $p$ and $q$ do
not need to be rationally related in order that the Einstein-Sasaki
metric can be complete.  

\section{A General Class of Solutions}

In this section, we consider a more general class of Einstein-Sasaki
metrics in dimension $D=d+1$, constructed as circle bundles over 
$d$-dimensional Einstein-K\"ahler spaces.  The $d$-dimensional 
Einstein-K\"ahler space is itself constructed as a complex line 
bundle over a product of $N$ Einstein-K\"ahler spaces, with dimensions
$n_i$ and metrics $d\Sigma_{n_i}^2$.  Thus $d=2 + \sum_{i=1}^N n_i$,
and the $d$-dimensional Einstein-K\"ahler metric will be written as
\be
ds_d^2 = dt^2 + c^2 \Big(d\tau' -\sum_{i=1}^N p_i A_\1^i
\Big)^2 + \sum_{i=1}^{N} a_i^2 d\Sigma_{n_i}^2\,,
\label{moregenmetric}
\ee
where $J_\2^i=dA_\1^i$ is the K\"ahler form for the Einstein K\"ahler
metric $d\Sigma_{n_i}^2$, with cosmological constant $\lambda_i$.
The metric (\ref{moregenmetric}) is Einstein K\"ahler with cosmological
constant $\Lambda$, provided that
the functions $c$ and $a_i$ satisfy the first-order equations
\be
\fft{\dot a_i}{a_i} = \fft{p_i\, c}{2a_i^2}\,,\qquad
\fft{\dot c}{c}= \fft{\lambda_1-\Lambda a_1^2}{p_1\,c} -
\ft12 \sum_{i=1}^N \fft{n_i\, \dot a_i}{a_i}\,,
\ee
together with the set of algebraic constraints
\be
\lambda_i\, p_j - \lambda_j\, p_i +
\Lambda(\lambda_j\, a_i^2 - \lambda_i\, a_j^2) = 0\,.
\ee
Note that there are $(N-1)$ independent constraints.
The solutions can be obtained straightforwardly, given by
\be
a_i^2 = p_i\, r + \ell_i\,,\qquad
c^2 = \fft{2}{\prod a_i^{n_{i}} }
\int_0^r \fft{\lambda_1-\Lambda a_1^2}{p}\,
\prod_i a_i^{n_i}\,,
\ee
where the coordinate $r$ is defined by $dr=c\, dt$.  The integration
constants $\ell_i$ satisfy the constraints
\be
\beta = \fft{\Lambda p_i-\lambda_i}{\Lambda\, p_i}=
{\rm constant}\,,\qquad\qquad \hbox{for all $i$}\,.
\label{betageneq}
\ee

     The $D=d+1=3+\sum n_{i}$ dimensional Einstein-Sasaki metric
is given by
\be ds^2_D=(d\psi' + 2 \cA_\1)^2 + ds_{d}^2\,,\label{genDmetric}
\ee
with $\cA_\1$ given by
\be
\cA_\1 = r d\tau' - \sum_{i=1}^N a_i^2 A_\1^i\,.
\ee
For the solution to be Einstein, we must have $\Lambda=4+\sum n_i$
(after choosing, without loss of generality, $\lambda_i=1$).

       To study the global structure of the metrics, it is appropriate
to make the coordinate transformation $\psi'=2\tau$ and $\tau'=\beta^{-1}
(\tau-\Lambda^{-1}\,\psi)$.  The metric becomes
\bea
ds_D^2 &=& \fft{dr^2}{c^2} + \fft{4c^2}{\Lambda^2(c^2 + 4(\beta+r)^2)}
(d\psi -\sum \lambda_i A_\1^i)^2 +
\sum a_i^2\, d\Sigma_{n_i}^2\nn\\
&&+\fft{c^2 + 4 (\beta +r)^2}{\beta^2} \Big(d\tau +
\sum \ell_i\,A_\1^i -
\fft{c^2 + 4r(\beta +r)}{\Lambda(c^2 + 4 (\beta +r)^2)}
(d\psi - \sum \lambda_i A_\1^i)\Big)^2\,.
\eea
As in the $D=7$ case we discussed in the previous section, the rate
of the collapsing of the circle parameterised by $\psi$ is the same
at all the roots of $c^2(r)=0$.  The period of $\psi$ must then be
$2\pi$.  Consideration of the connection on the
fibres parameterised by $\tau$ (which never shrink to zero) 
implies the conditions
\bea
&&\fft{\ell_i}{\ell_{\sst N}} =\alpha_i \equiv \hbox{rational number}
\,,\qquad\qquad i=1,2\,\cdots,N-1\,,\nn\\
&&\fft{r_+}{\Lambda\,(\beta+r_+)\,\ell_{\sst N}}=\gamma \equiv
\hbox{rational number}\,.\label{gencon}
\eea
Note that the $p_i$ do not have to be rationally related, but they
satisfy the conditions (\ref{betageneq}).  Substituting (\ref{gencon})
and (\ref{betageneq}) to $c^2=0$ equation, we obtain a polynomial
equation in $\ell_{\sst N}$ of order $1 + \ft12 \sum n_i$, with
rational coefficients that are polynomials in $\alpha_i$ and $\gamma$.
Without loss of generality, we can choose $0\le p_1\le p_2\le
\cdots\le p_{\sst N}$ and $\lambda_i=1$.  The constant $\ell_{\sst N}$
must lie in the range $0<\ell_{\sst N}< \Lambda^{-1}$.  Thus provided
$\ell_{\sst N}$ satisfies this condition, the corresponding set of 
rational numbers $(\alpha_i,\gamma)$ gives a non-singular
Einstein-Sasaki metric.

     It is also possible to find special solutions where $\ell_{\sst
N}$ is rational too.  These correspond to cases where the parameters
$p_i$ are all relatively-prime integers (after rescaling).  Such
solutions occur sporadically, and their significance is unclear.

\section{Conclusions}

In this note, we obtained an infinite number of Einstein-Sasaki
metrics in $D=2n+3$ dimensions, which are circle bundles over
Einstein-K\"ahler $(2n+2)$-spaces. These spaces are themselves complex
line bundles over a product of $N$ Einstein-K\"ahler manifolds of
diverse dimensions $n_i$.  Locally, the Einstein-Sasaki metrics are
characterised by $N$ real parameters.  
Global considerations for non-singular metrics that extend
smoothly onto complete compact manifolds restrict these $N$ parameters
to be rational within a certain region.

  We focused our attention principally on seven-dimensional examples,
which provide natural supersymmetric compactifying manifolds for M-theory.

\section*{Acknowledgment}

J.F.V.P. thanks the Princeton Institute for Advanced Study for hospitality
during the course of this work.

\section*{Note Added}

   After this work was completed, a paper appeared that also obtained 
the local form of the Einstein-Sasaki metrics that we have constructed in this
paper \cite{gaunt3}.

\end{document}